\documentstyle[12pt,a4,epsf]{article}
\begin{document}

\begin{center}
{\Large \bf   The influence of shell effects on the fissility
of neutron deficient uranium  isotopes at \hbox{N $\simeq$ 126}.}
\vspace{5mm} \\
{\normalsize  M.~Veselsk\' y~$^a$, D.D.~Bogdanov, A.N. Andreyev, V.I. Chepigin,\\
A.P. Kabachenko, O.N. Malyshev, J. Roh\' a\v c, R.N. Sagaidak, \v S. \v S\' aro$^a$,\\
G.M. Ter-Akopian,  A.G. Popeko, A.V. Yeremin}\\
\vspace{2mm}
{\normalsize \it  Flerov Laboratory of Nuclear Reactions, JINR, 141 980
Dubna, Russia} \\
{\normalsize \it $^a$ Department of Nuclear Physics,  Comenius University,
SK-842 15 Bratislava, Slovakia} \\
\end{center}

\begin{abstract}
\noindent
Production cross sections and excitation functions of neutron deficient
isotopes of U, Pa and Th have been measured in the reaction of
$^{22}Ne + ^{208}Pb$. The comparison of experimental production cross sections
of uranium isotopes with the neutron number \hbox{$126 \le N \le 134$}
to the values calculated using statistical model of nuclear deexcitation
has been carried out. There is shown, that none of the commonly used
methods of introduction of shell effect influence on the production
cross sections of evaporations residues allows to reproduce
unambigiously used set of experimental data. Simple semiempirical
method of introduction of shell effect influence in fission channel
is proposed and good agreement with production cross sections of
neutron deficient isotopes in the region from Bi to U is reached.
\end{abstract}

\section*{{\large{\bf Introduction}}}

   The investigated production cross sections of Th - U isotopes with  N
$\simeq$ 126 are interesting for several reasons. At first, the shell
correction energy to the ground state of these isotopes is large and
comparable to their liquid drop fission barrier.  This is a good reason to
assume that the comparison of experimental cross sections for xn-, pxn-
and $\alpha$xn-reactions  with  calculated values, obtained using the
statistical  model of compound nucleus deexcitation, will allow to
determine the role and influence of shell effects on the fission
probability of an excited compound nucleus and, consequently,
to determine the values
of production cross sections of different evaporation products. Correct shell
correction values are very important for transfermium nuclei.  On the
other side the available  experimental data allow  model dependent
interpretations based on  different physical principles.
We measured the production cross sections of evaporation residues
produced in xn-, pxn- and $\alpha$xn-reaction channels
bombarding $^{208}$Pb targets
with $^{22}$Ne ions having energies from 110 MeV to 155 MeV.
The data
obtained in these experiments, together with the data on cross sections
of neutron deficient isotopes of U, Pa, Th, from the reactions of
$^{20,22}Ne + ^{208}Pb$ and $^{27}Al + ^{197}Au$ \cite{AN5,AN6,AN7,AN8},
allow to follow in details the changes in production cross sections of nuclei
in the Th-U region at the variation of neutron  numbers  from 134 to 124.
The analysis of experimental data and the  comparison with the statistical
model of compound nucleus deexcitation is the aim of this article.\\

\section*{{\large{\bf Experimental methodics and results of measurements}}}

The experiments have been performed with the use of the external beam
of the cyclotron U400 at JINR Dubna. The  $^{22}Ne$ beam  with initial
energies of 130 MeV and 160 MeV had a mean intensity of $2 \times 10^{11}
s^{-1}$. The energy of the beam was varied by steps of 3-6 MeV using
aluminium and titanium  degraders. The energy of the projectiles was
determined by measuring the energy of ions scattered on a $250\mu g/cm^2$
gold foil and on the target foil to 30$^\circ$ using Si-detectors,
placed  in front and  behind the target foil.  As target enriched
$^{208}Pb$(99\%) was used having a thickness of $400 \pm 150 \mu g/cm^2$
evaporated onto a 66 $\mu m$ Al foil.\\
The complete fusion reaction products were separated from  deep inelastic
transfer products and from projectiles with the kinematic separator
VASSILISSA \cite{YE1}. This three-stage electrostatic separator have an
angel of acceptance of 15 mstr and an energy acceptance of $\pm 10\%$
of electric rigidity. At a transit time about 1 $\mu s$ the separator is
able to separate very effectively complete fusion reaction products, transfer
reaction products and projectiles.
	The evaporation residues and their $\alpha$- decay energies
were detected and measured in the separator's focal plane with a detector
system \cite{AN9} consisting of two time-of-flight detectors (0.5 ns
time resolution) and eight strip Si-detector array (60 x 60 mm$^2$,
$\approx$ 15 keV energy resolution for 6 - 9 MeV $\alpha$-particles).
For fine calibration of the Si-detector strips implanted $\alpha$-decay
products of the reactions $^{22}$Ne + W, Os, Pt were used. The electronic
device allowed to register all events, related to the evaporation residues
and their $\alpha$-decay products, including their time consequence. The
time - amplitude correlation analysis of the whole set of registered
events allowed to determine genetically related $\alpha$-decay chains
and to identify the initial nucleus. From the time distribution of
$ER- \alpha_{1}- \alpha_{2}$ correlation it was possible to determine
the half-life of the daughter nucleus.\\
The statistical accuracy of yield determinations in our experiments
was $\pm (15 -20)\%$ for the uranium and protactinium isotopes and $\pm (5 -
10)\%$ for the thorium isotopes. Taking into consideration all the
parameters having influence, the absolute cross section values for the
evaporation residues were in our experiments determined to be $\pm 50\%$.
The accuracy of relative cross section measurements were two - three times
better.\\
        The excitation energies of compound nuclei were calculated on the basis
of experimental mass values of nuclei, taken from \cite{WA1}. The cross
section values for the isotopes of U, Pa and Th, measured in the excitation
energy region from 30 MeV to 80 MeV of the $^{230}U$ compound nucleus,
are given in tab. 1.

{\footnotesize
\begin{center}
 Table 1. Cross sections for evaporation  products in
 $^{22}$Ne+$^{208}$Pb reaction.  \\
\vskip 0.4cm
\begin{tabular}{|r|c|c|c|c|c|c|c|c|c|c|c|c|c|c|}
\hline
E$_{Lab}$ & E$^{*}$ & \multicolumn{4}{|c|}{xn-channels, $\mu$b} & \multicolumn{3}{|c|}{pxn-channels, $\mu$b} & \multicolumn{6}{|c|}{$\alpha$xn-channels, $\mu$b}\\ \hline
MeV  & MeV& 4n   &   5n  &  6n  & 7n   & p5n & p6n & p7n & $\alpha$2n & $\alpha$3n & $\alpha$4n & $\alpha$5n & $\alpha$6n & $\alpha$7n \\ \hline
101  & 31 & 0.7  &       &      &      &     &     &     & 210       &  140      &            &            &           &          \\
109  & 38 & 6.0  &       &      &      &     &     &     & 310       &  330      &            &            &           &          \\
112  & 41 & 3.1  &  0.5  &      &      &     &     &     &  60       &  380      &   50       &            &           &          \\
116  & 45 & 0.8  &  1.9  &      &      &     &     &     &  40       &  310      &   90       &            &           &          \\
122  & 50 & 0.2  &  1.8  & 0.3  &      &     &     &     &  20       &  120      &  230       &     50     &           &          \\
130  & 57 &      &  0.9  & 0.9  &      & 1.4 & 0.1 &     &           &   30      &  250       &    250     &     20    &          \\
137  & 64 &      &  0.4  & 0.2  & 0.1  & 3.7 & 0.3 &     &           &   10      &   60       &    310     &     50    &          \\
142  & 68 &      &       & 0.1  & 0.3  & 1.1 & 1.9 & 0.4 &           &           &   40       &    280     &    120    &    10    \\
148  & 74 &      &       &      & 0.2  & 0.9 & 2.6 & 0.4 &           &           &   20       &    120     &    200    &    30    \\
153  & 78 &      &       &      & 0.1  & 0.7 & 2.2 & 0.8 &           &           &            &     60     &    140    &    80    \\ \hline
\end{tabular}
\end{center}
}

\section*{{\large{\bf The comparison of experimental data and statistical calculations,
conclusions}}}

        For the analysis of experimental data the HIVAP code \cite{RE1}
was used, where the production cross sections of evaporation residues,
created in complete fusion reactions , are calculated in the framework
of the statistical model of compound nucleus deexcitation. The level
densities were calculated using the well known equations of the Fermi gas
model (neglecting the effects of collective amplification) taking  the
shell effects in level densities  into consideration in a fenomenological
way, according to Ignatyuk \cite{IG3}.

\begin{equation}
a_{\nu}(E^{*}) = \tilde a_{\nu}(1+(1 - exp(-E^{*}/D))\Delta W_{\nu}(A,Z)/
E^{*})
\end{equation}
where $E^{*}$ - the compound nucleus excitation energy, D = 18.5 MeV - the
shell effects attenuation, and $\Delta W_{\nu}(A,Z)$ - the shell correction to
the nucleus mass after the evaporation of particle $\nu$  (neutron,
proton, or $\alpha$ -particle). We considered the level density parameter
$\tilde a_{f}$ to be equal to the asymptotic value of the level density
in the channel of $\tilde a_{\nu}$ particles evaporation and to be independent
of the excitation energy. In this way  the ratio of asymptotic values
of the level density parameter in the fission and evaporation channels -
$\tilde a_{f}/\tilde a_{\nu}$ were  in the calculations considered to be equal
one. The experimental arguments for such a choice of the
$\tilde a_{f}/\tilde a_{\nu}$  value were discussed in details in an earlier
work \cite{AN2}.
	The full fission barrier was calculated as the sum of the
liquid drop and shell effect components.

\begin{equation}
B_{f}(l) = C B_{f}^{LD}(l) + \Delta B_{f}^{Shell}.
\end{equation}
	The liquid drop component of the fission barrier  ($B_{f}^{LD}$)
was calculated after the charged liquid drop model of Cohen-Plasil-Swiatecki
\cite{CO1}. The shell component of the fission barrier ($\Delta B_{f}^{Shell}$)
was taken as the difference between the liquid drop \cite{MY1} and
measured \cite{WA1} mass of the nucleus, i.e. as the module of
$\Delta W_{\nu}(A,Z)$. Coefficient C in the liquid drop component of the
fission barrier was used to achieve full agreement.\\
	 As can be seen from the experimental data, given in tab. 1, the
$\alpha$xn reaction channel is dominant. Even at relatively low
excitation energies ($\approx$ 35 MeV) the cross section of the
$\alpha$xn-channel is two orders of magnitude higher than for
the xn- and pxn-channel.
With increasing excitation energy the difference in cross sections reaches
three orders of magnitude as the consequence of the fast drop of the cross
section for the xn-channel. Such a large difference in cross sections makes
it very difficult to describe these reaction channels using only one set of
model parameters. We believe, that for the investigated types of reactions
for correct calculations of cross sections it is not enough to make careful
calculations of the fission process, but because of an important role of the
charged particle evaporation, this process have to be also correctly
calculated.\\
        The experimental and calculated cross section values were compared
for the reactions of $^{22}Ne + ^{208}Pb$ and  $^{27}Al + ^{197}Au$.
From  the three deexcitation channels, xn, pxn, and $\alpha$xn, the
first one is the most important. This channel undergoes the most radical
changes in cross sections and therefore is less sensitive to errors in
cross section measurements. In Fig. 1 the experimental cross sections for
neutron deficient isotopes of uranium, created in xn-reactions with light
ions (A $\leq$ 40) are shown.  The figure contains data from the present
work and also  from ref \cite{AN5,AN6,AN7,AN8}, where cross section values
for the reactions $^{22}Ne + ^{208}Pb$ and $^{27}Al + ^{197}Au$ are
presented. Points denote experimental values at maximum yields, full
lines denotes values of HIVAP calculations.
%
        The dash-dotted line in Fig. 1 ( line 1 ) denotes the results of
calculations,
made using the standard approach, described above in this paragraph. In
this calculations only one parameter - the factor C = 0.65 was used.
This value of C is typical for the liquid drop fission barrier of nuclei
for the region of Pb -U \cite{AN2,AN3}. These calculations give good agreement
with experimental values only for isotopes with $130 \leq N \leq 134$. The
calculated cross sections for $^{218,219}U$ with N = 126, 127 are 10 -20
times higher than the experimental ones.  Additional calculations have
showed that for correct cross sections for these isotopes coefficient C should be
as low as 0.45 (line 3 in Fig. 1). But in this case the cross section
values for heavier isotopes of uranium are running far down.\\
\begin{figure}[h]
\vbox{
\vspace*{-1.0cm}
\epsfxsize=10cm \epsfbox{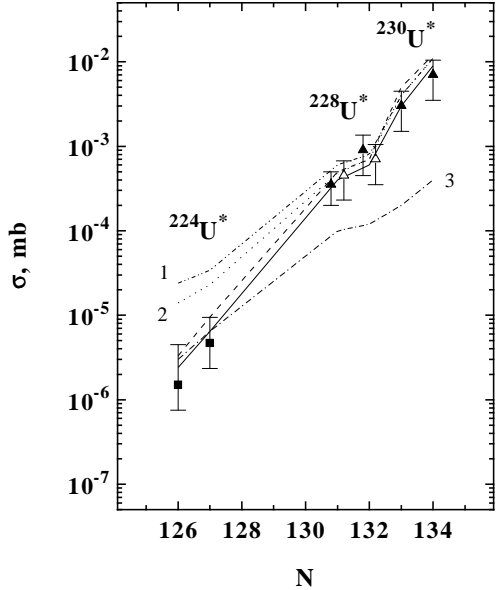}
\vspace*{-2cm}
}
\caption{
Experimental and calculated values of maximal production cross
sections of xn-reactions. For details see text.
}
\label{figuran1}
\end{figure}
\noindent
        We also tested  the sensitivity of the calculations to systematic
errors in the determination of shell effects corrections to the ground
state. For this purpose the shell effect corrections in formulas for the
level density and fission barrier for all the nuclei of the evaporation
cascade  were at the same time increased or decreased by $30\%$. In absolute
measures such a variation is equivalent to shell correction of $\pm 0.8$
MeV for the end nucleus of the evaporation cascade. The result of calculations
for the decreased value of shell correction is shown in Fig. 1 (line 2).
As we can see, even such a strong change of shell correction leads only
to cross section deviation by a factor of 2 - 3 and do not give an
explanation for the very low cross section values for $^{218,219}U$.
This is understandable because the role of shell correction in level densities
leads to the decrease of the evaporation width and in the fission
barrier to the decrease of fission width. Therefore simultaneous change
of the shell effect corrections in the evaporation and fission channels
has a consequence of strong  mutual elimination of both effects in cross
section calculations.\\
	Very good agreement between experimental and calculated cross section
values can be achieved using the approach, recommended in refs. \cite{SCH1,SA1}
and the decrease of parameter D for the level density calculation in the
particle evaporation channel. The results of such calculations, with
D = 10.5 MeV and C = 0.65 are shown in Fig. 1 as a dashed line. As in the
figure shown, at such a choice of parameters it is possible to describe
equally well the production cross section of isotopes with both, considerable
and negligible shell corrections, but this approach does not seem to
have unambigious physical meaning.
  There is the another, very simple variant how to take into account the
shell effects in the fission channel for the production cross section
of evaporation products. The full line in Fig. 1 is the result of
calculations, where with the free parameter C = 0.65  not only the
liquid drop component but all the formula  was multiplied.

\begin{equation}
B_{f}(l) = C ( B_{f}^{LD}(l) + \Delta B_{f}^{Shell} ).
\end{equation}
As we can see, this variant of calculations
gives the best agreement with the experimental data for the xn-channel
through all the investigated region of nuclei.
To test the universality of this new approach to the problem we
applied it to cross section calculations of evaporation nuclei, experimentally
investigated  in work \cite{AN3}. The investigated set
contains data for about 50 product nuclei, created in more that 15 reactions.
The results of calculations are presented in figures 2a and 2b, where
the optimum values of C are given, obtained from the comparison of
experimental and calculated cross section values. Fig. 2a shows the results
of calculations, where the free parameter C was applied to the whole
fission barrier. The results shown in Fig. 2b, taken from ref. \cite{AN3},
are completed with data for the uranium isotopes, and represent the
calculations where the shell effects are accounted in the standard way.
The comparison of both figures  shows that the new approach to account
the shell effects in the fission channel describes better the set
of analyzed data. When applying this approach, there is no need for the
variation of C to get correct description of production cross sections
for nuclei with $N \approx 126$. This approach gives also reasonable
reproduction of production cross sections of Th isotopes produced
in $\alpha$xn- channels.\\
        According to our opinion the mathematical simplicity of the new
approach to the production cross section calculations of evaporation nuclei
does not mean the simplicity of the investigated  physical process itself.
The obtained results indicate some limitations for the use of the given
set of experimental data to try explain the presence of several
physical effects in the same process and  even from the experimental data
to try to determine the value of their parameters.\\
 	The authors would like to express their gratitude  to
prof. Yu.Ts. Oganessian for his support of this work, to Yu.A. Muzychka and
B.I. Pustylnik for useful discussions and remarks and to A.V. Taranenko
for his help at the experiments.
\begin{figure}[h]
\hspace*{2cm}
\vbox{
\epsfysize=10cm \epsfbox{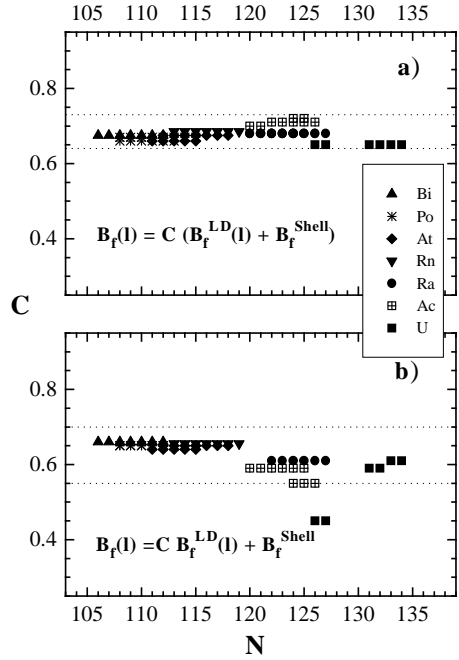}
}
\caption{
Optimum values of parameter C necessary for correct calculation
of production cross section of neutron deficient isotopes in the
region from Bi to U in the reactions with the heavy ions with A $\le$ 40.
(a) Parameter C used to scale the whole fission barrier.
(b) Parameter C used to scale the macroscopic part of fission barrier.}

\label{figuran4}
\end{figure}

\newpage

\begin{figure}[htb]
\vbox{
\epsfxsize=10cm \epsfbox{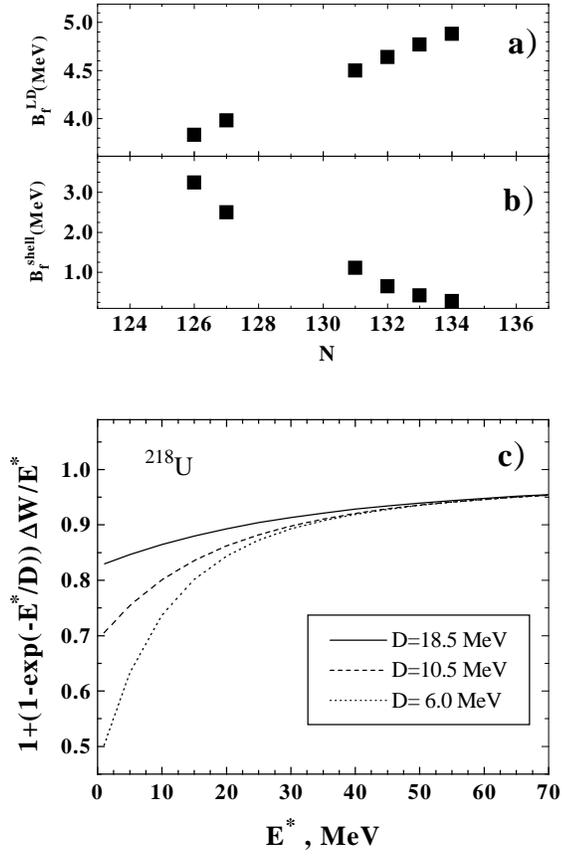}
}
\caption{
Values of liquid drop components (a) and shell components (b)
of fission barriers of uranium isotopes. Dependence of level density
parameter on excitation energy for three values od parameter D (c):
D = 18.5, 10.5 and 6.0 MeV.
}
\label{figuran2}
\end{figure}

\begin{figure}[htb]
\vbox{
\epsfxsize=10cm \epsfbox{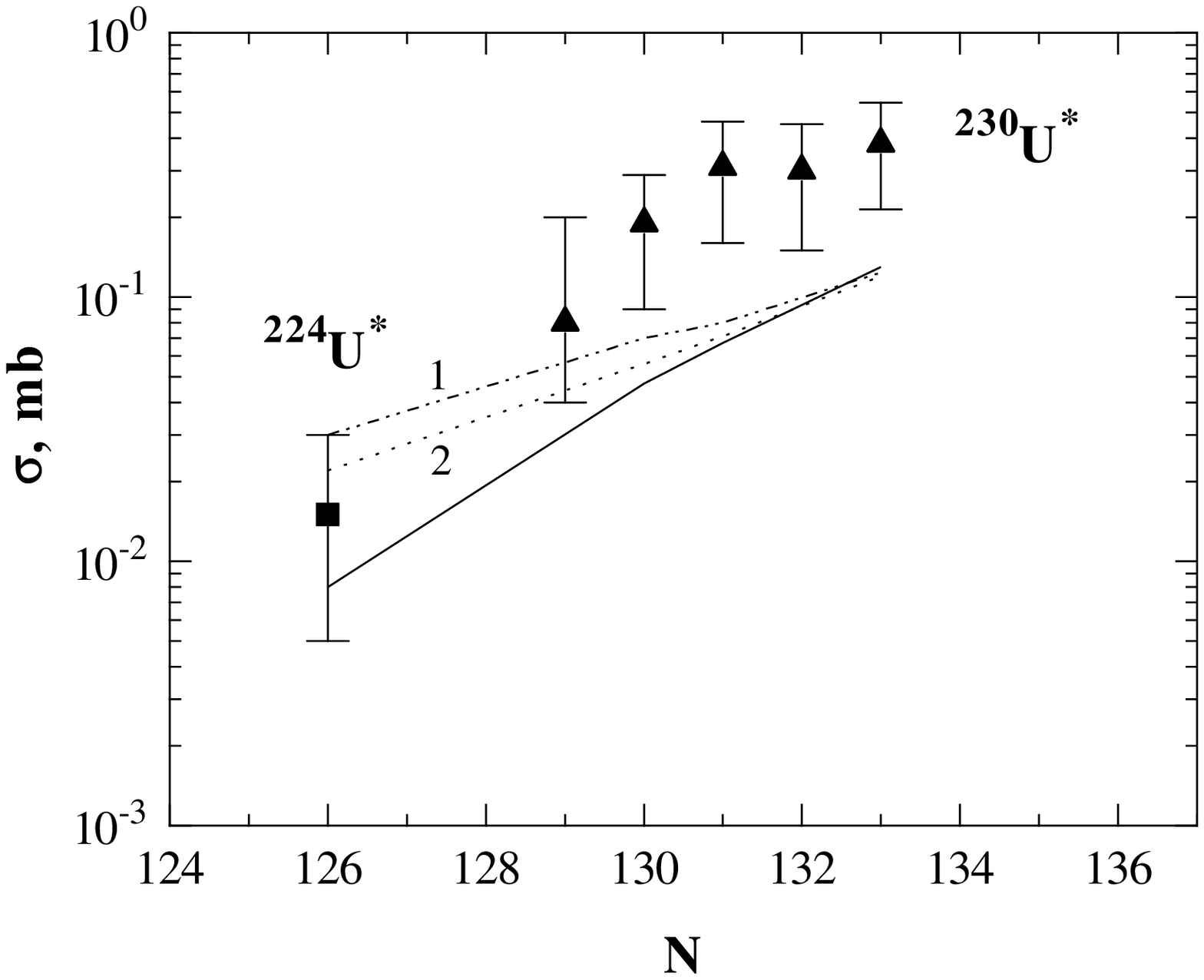}
}
\caption{
Experimental and calculated values of maximal production cross
sections of $\alpha$xn-reactions. Symbols and lines as in Fig. \ref{figuran4}.
}
\label{figuran3}
\end{figure}


\noindent

\footnotesize

\begin{thebibliography}{99}
\bibitem{AN5}
A.N. Andreyev, D.D. Bogdanov, V.I. Chepigin, A.P. Kabachenko,
O.A. Orlova,  G.M. Ter-Akopian and A.V. Yeremin, Yadernaya Fizika 50 (1989)
619.
\bibitem{AN6}
A.N. Andreyev, D.D. Bogdanov, V.I. Chepigin,  A.P. Kabachenko,
O.N. Malyshev,  G.M. Ter-Akopian and A.V. Yeremin, Yadernaya Fizika 53 (1991)
895.
\bibitem{AN7}
A.N. Andreyev, D.D. Bogdanov, V.I. Chepigin, A.P. Kabachenko,
O.N. Malyshev, R.N. Sagaidak,  G.M. Ter-Akopian and A.V. Yeremin,
Z. Phys. A 342 (1992) 123.
\bibitem{AN8}
A.N. Andreyev, D.D. Bogdanov, V.I. Chepigin,  A.P. Kabachenko,
O.N. Malyshev, R.N. Sagaidak,  G.M. Ter-Akopian, M. Veselsky and
A.V. Yeremin, Z. Phys. A 345 (1993) 247.
\bibitem{YE1}
A.V. Yeremin, A.N. Andreyev, D.D. Bogdanov, V.I. Chepigin, V.A. Gorshkov,
A.P. Kabachenko, A.Yu. Lavrentjev, O.N. Malyshev, A.G. Popeko,
R.N. Sagaidak,  S. Sharo, G.M. Ter-Akopian, E.N. Voronkov, A.V. Taranenko,
Nucl. Instr. Methods A 350 (1994) 608.
\bibitem{AN9}
A.N. Andreyev, V.V. Bashevoy, D.D. Bogdanov, V.I. Chepigin,  A.P. Kabachenko,
O.N. Malyshev, J. Rohac, S. Saro, A.V. Taranenko,  G.M. Ter-Akopian,
A.V. Yeremin,
Nucl. Instr. Methods A 364 (1995) 342.
\bibitem{WA1}
A.H. Wapstra, G. Audi, R. Hoekstra,
Atomic Data and Nuclear Data Tables, Vol. 39, 1989, p. 281.
\bibitem{RE1}
W. Reisdorf, Z. Phys. A 300 (1981) 227.
\bibitem{IG3}
A.V. Ignatyuk, G.N. Smirenkin, A.S. Tishin, Yadernaya Fizika 21 (1975) 485.
\bibitem{AN2}
A.N. Andreyev, D.D. Bogdanov, V.I. Chepigin, A.P. Kabachenko, O.N. Malyshev,
Yu.A. Muzychka, Yu.Ts. Oganessian, A.G. Popeko, B.I. Pustylnik, R.N. Sagaidak,
A.V. Taranenko, G.M. Ter-Akopian and A.V. Yeremin, JINR Rapid Communications
4[72]-95, 43.
\bibitem{CO1}
S Cohen, F. Plasil, W.J. Swiatecki, Ann. Phys. (N.Y.) 82 (1974) 557.
\bibitem{MY1}
W.D. Myers, W.J. Swiatecki, Ark. Fyz. 36 (1967) 343.
\bibitem{AN3}
A.N. Andreyev, D.D. Bogdanov, V.I. Chepigin, A.P. Kabachenko, O.N. Malyshev,
Yu.Ts. Oganessian, A.G. Popeko, J. Rohac, R.N. Sagaidak, S. Saro,
G.M. Ter-Akopian, M. Veselsky and A.V. Yeremin, JINR Rapid Communications
5[73]-95, 57.
\bibitem{SCH1}
K.-H. Schmidt, W. Faust, G. M\" unzenberg, W. Reisdorf, H.-G. Clerc,
D. Vermeulen and W. Lang, Phys. and Chem. Fission, 1979, IAEA Vienna,
Vol. 1, p. 409-420.
\bibitem{SA1}
C.-C. Sahm, H.-G. Clerc, K.-H. Schmidt, W. Reisdorf, P. Armbruster,
F.P. He\ss berger, J.G. Keller, G. M\" unzenberg and D. Vermeulen,
Nucl. Phys. A 441 (1985) 316.
\bibitem{IG2}
A.V. Ignatyuk, K.K. Istekov, G.N. Smirenkin, Yadernaya Fizika 37
(1983) 316.
\end{thebibliography}
\end{document}